\documentclass[aps,prb,twocolumn,superscriptaddress, showpacs]{revtex4}

\def\schaal{1}

\usepackage{graphicx}
\usepackage{times}

\begin{document}

\title{In search for the superconducting spin-switch: Magnetization induced resistance switching effects in  La$_{0.67}$Sr$_{0.33}$MnO$_3$/YBa$_2$Cu$_3$O$_{7-\delta}$ bi- and trilayers}

\author{M. van Zalk}
\email[]{m.vanzalk@utwente.nl}
\author{M. Veldhorst}
\author{A. Brinkman}
\affiliation{Faculty of Science and Technology and MESA$^+$ Institute for Nanotechnology, University of Twente, 7500 AE Enschede, The Netherlands}
\author{J. Aarts}
\affiliation{Kamerlingh Onnes Laboratory, Leiden University, 2300 RA Leiden, The Netherlands}
\author{H. Hilgenkamp}
\affiliation{Faculty of Science and Technology and MESA$^+$ Institute for Nanotechnology, University of Twente, 7500 AE Enschede, The Netherlands}

\date{\today}

\begin{abstract}
We have studied the influence of the magnetization on the superconducting transition temperature ($T_c$) in bi- and trilayers consisting of the half-metallic ferromagnet La$_{0.67}$Sr$_{0.33}$MnO$_3$ (LSMO) and the high-temperature superconductor YBa$_2$Cu$_3$O$_{7-\delta}$ (YBCO). We have made use of tilted epitaxial growth in order to achieve contacts between the two materials that are partly in the crystallographic $ab$-plane of the YBCO. As a result of uniaxial magnetic anisotropy in the tilted structures, we observe sharp magnetization switching behavior. At temperatures close to $T_c$, the magnetization switching induces resistance jumps in trilayers, resulting in a magnetization dependence of $T_c$. In bilayers, this switching effect can be observed as well, provided that the interface to the ferromagnetic layer is considerably rough. Our results indicate that the switching behavior arises from magnetic stray fields from the ferromagnetic layers that penetrate into the superconductor. A simple model describes the observed behavior well. We find no evidence that the switching behavior is caused by a so-called superconducting spin-switch, nor by accumulation of spin-polarized electrons. Observation of magnetic coupling of the ferromagnetic layers, through the superconductor, supports the idea of field induced resistance switching.
\end{abstract}


\maketitle

\section{Introduction}

The interplay between superconductivity and ferromagnetism is a rapidly developing field in condensed matter physics. In hybrid heterostructures, where the two different orders meet at the interface, interesting physics arises. One of the promising structures is the so-called superconducting spin-switch, \cite{tagirov1999lfs, buzdin1999sod} which consists of two ferromagnetic metallic layers, sandwiching a superconductor. An early theoretical proposal for a spin-switch, involving ferromagnetic insulators was made by De Gennes. \cite{degennescbf} Here, the average exchange field induced in the superconductor depends on the relative orientation of the ferromagnetic layers. As a result, the superconducting transition temperature, $T_c$, depends on this orientation. Recently, such geometries were investigated for the case of metallic weak ferromagnets and it was predicted that, under the right circumstances, superconductivity can be switched on and off by applying a small external field. \cite{tagirov1999lfs, buzdin1999sod} This switching was suggested to result from interference between the superconducting wave functions transmitted through the S/F interface and reflected at the F surface.  An alternative scenario for spin-switching is in terms of crossed Andreev reflection: \cite{giazotto2006car} when the ferromagnetic layers are magnetized in the antiparallel direction, Cooper pair formation due to crossed Andreev reflection is enhanced, compared to the parallel configuration. This effect is the largest for strongly spin-polarized magnets, when crossed Andreev reflection occurs \emph{only} in the case of antiparallel magnetization.

Although full switching of superconductivity has never been observed, a resistance drop has been found in F/S/F systems with weak ferromagnets when switching the magnetization from the parallel (P) to the antiparallel (AP) state. \cite{gu2002mod,potenza2005sfc} In systems with strong ferromagnets, the opposite effect has been observed by Rusanov \emph{et al.} \cite{rusanov2006iss}, which was attributed to an increased number of quasiparticles in the superconductor as a result of the enhanced reflection of the spin-polarized quasiparticles. However, Moraru \emph{et al.}~found the standard spin-switch effect in a comparable system. \cite{moraru2006mdt, moraru2006oss} The contradictory results might be related to the employment of the exchange bias mechanisms in some of these works. \cite{stamopoulos2007ese} Recently, $T_c$ shifts in F/I/S/I/F multilayer systems were observed that could not be fully explained by the spin-switch effect, but were partly attributed to spin imbalance in the superconductor, induced by the ferromagnet. \cite{miao2007isp} However, it was pointed out by Steiner \emph{et al.} \cite{steiner2006mso} that stray fields due to specific magnetic domain configurations can lead to changes in $T_c$. Stamopoulos \emph{et al.} reported stray-fields-based magnetoresistance in Ni$_{80}$Fe$_{20}$/Nb/Ni$_{80}$Fe$_{20}$ trilayers, that emerge from a magnetostatic coupling of the ferromagnetic layers. \cite{stamopoulos2007sfb,stamopoulos2007seb} The importance of stray fields was further established by Carapella \emph{et al.}, who found that a glassy vortex phase induced by magnetic stray fields explains the switching behavior in their Co/Nb/Co trilayers. \cite{carapella2008lft} Thus, magnetic stray field effects are a potential problem for the interpretation of data obtained on structures with ferromagnets in close proximity to superconductors.

Studies on F/S hybrid systems have not been limited to conventional superconductors and ferromagnets. Combinations of the oxide materials La$_{0.67}$Sr$_{0.33}$MnO$_3$ (LSMO) and La$_{1-x}$Ca${_x}$MnO$_3$ (LCMO) with YBa$_2$Cu$_3$O$_{7-\delta}$ (YBCO) have been used because of the high spin-polarization of LSMO \cite{park1998deh} and the good lattice match, allowing the growth of epitaxial structures. In these systems, large magnetoresistance and an inverse spin-switch effect were found and attributed to the transmission of spin-polarized carriers into the superconductor. \cite{pena2005gmf,nemes2008ois} Vortex effects were ruled out as a cause for the observed phenomena, since no effects were seen in bilayers. Anisotropic magnetoresistance effects were excluded on the basis of the absence of a dependence of the magnetoresistance peak on the relative orientation of current and magnetic field. \cite{visani2007sdm} However, the role of spin injection in LCMO/YBCO structures is not entirely clear. Gim \emph{et al.} \cite {gim2001cii} found no conclusive evidence for suppression of superconductivity from their quasiparticle injection experiments using LCMO/LSMO and YBCO. A similar conclusion was reached recently by Deng \emph{et al.} \cite{deng2008ssp} from mutual inductance measurements on YBCO/LCMO bilayers, which were optimized for the experiment by growing YBCO with the $c$-axis in the plane of the film. These kind of experiments are performed under equilibrium conditions in the bilayers and might be more comparable to the current-in-plane measurements of ref. \onlinecite{pena2005gmf} than quasiparticle injection experiments. In the mutual induction experiments, suppression of superconductivity was found near the coercive field of the LCMO layer, which was attributed to magnetic field effects. 

It has been known from other systems as well that the effects of field can be important; for example they can give rise to domain-wall guided superconductivity \cite{yang2004dws} and flux-flow induced giant magnetoresistance effects. \cite{bell2006ffi} The volume magnetization of LSMO $\mu_0 M$ can reach 0.8~T and it therefore is reasonable to expect a strong influence of stray fields. In a recent publication, Mandal \emph{et al.} \cite{mandal:094502} point at a distinct contribution of the dipolar field to the magnetoresistance in F/S/F trilayers with Y$_{0.6}$Pr$_{0.4}$Ba$_{2}$Cu$_3$O$_7$ used for the superconductor. However the relative contribution to the magnetoresistance of the depairing due to accumulation of spin-polarized electrons remains unclear. Furthermore, the higher resistance seen in the state of AP magnetization is not understood. 

\begin{figure}
\includegraphics[width=.5\columnwidth]{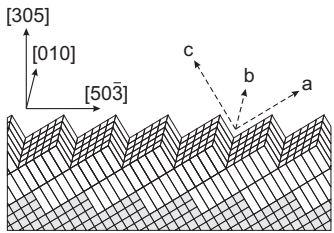}
\caption{\label{structure} Schematic picture of YBCO grown on STO (305). Indicated are the in-plane and out-of-plane crystallographic orientations and the YBCO $a$, $b$ and $c$-directions. The $c$-axis makes an angle of 31$^\circ$ with respect to the sample surface.}
\end{figure}

So far, $c$-axis oriented YBCO/LSMO superlattices, such as grown on SrTiO$_3$ (STO) (001) substrates, have been widely exploited. A disadvantage of these structures is the weak coupling between the superconductor and the ferromagnet, due to the strongly anisotropic nature of superconductivity in YBCO. In order to achieve coupling that is (partly) in the $ab$-plane, we will exploit coherently tilted epitaxial growth \cite{aarnink1992ssc} of YBCO on STO (305) substrates. On these substrates, YBCO grows with the $c$-axis making a 31$^\circ$ angle with respect to the sample surface, as indicated in Fig.~\ref{structure}. A second advantage of using the (305) oriented structures is that remarkably sharp magnetization switching behavior can be realized, caused by the induced uniaxial magnetic anisotropy, with the easy axis along the [010] direction. This enables us to prepare a well-defined state of parallel (P) or antiparallel (AP) magnetization in trilayers. 

In this article, we show that the trilayer resistance shows a sharp drop when the magnetization is switched from the AP to the P state within the superconducting transition. However, we find that the observed switching behavior is incompatible with the superconducting spin-switch model and models based on spin imbalance. We find a natural explanation in terms of stray fields from the LSMO layers that penetrate the superconductor. Our measurements show clearly that the switching behavior can be understood completely from changes in the effective field when one of the ferromagnetic layers switches. We will show that we can even obtain switching behavior in bilayers, as is expected within our model, by exploiting the controllable surface roughness of the ferromagnetic layers.

\section{Experimental details \label{experimental}}

\subsection{Film growth and characterization}

\begin{figure}
\includegraphics[scale=\schaal]{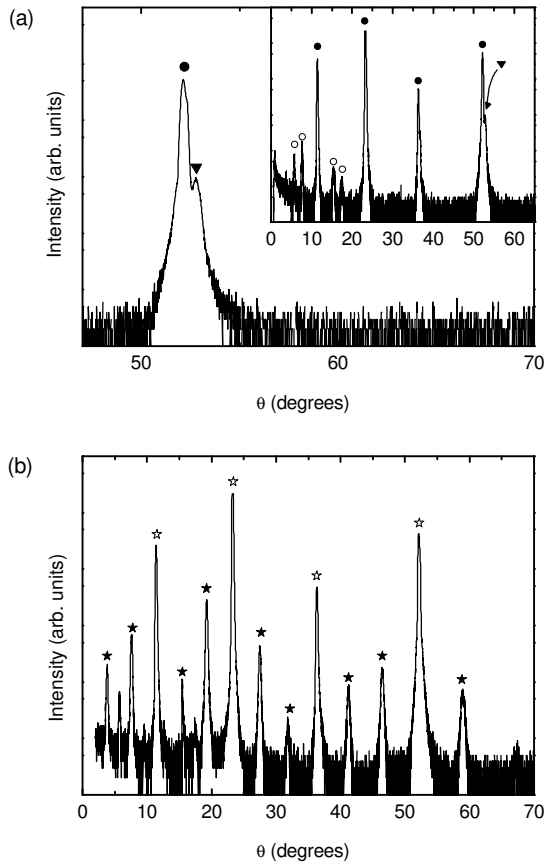}
\caption{\label{XRD} (a) $\theta$-$2\theta$ scan of LSMO grown on STO (305). Triangles denote LSMO peaks, which largely overlap the STO peaks, indicated by closed circles. Peaks indicated by open circles are due to higher harmonics in the beam. (b) $\theta$-$2\theta$ scan for a YBCO/LSMO bilayer. Filled stars correspond to YBCO peaks, open stars indicate overlapping STO, LSMO and YBCO peaks.}
\end{figure}

All thin films were grown on SrTiO$_3$ (STO) substrates. The STO~(001) substrates were chemically treated \cite{koster1998qis} and annealed for at least two hours at 950 $^\circ$C in an oxygen flow to produce atomically flat, TiO$_2$-terminated surfaces. For the (305) oriented substrates a single termination does not exist, but the surfaces were atomically flat and substrate steps were observed, due to a small miscut with respect to the (305) plane. The thin film heterostructures were grown with pulsed laser deposition using a laser fluency of 1.5~Jcm$^{-2}$ for both YBCO and LSMO. Film thicknesses were in the range of 50 to 150~nm for LSMO and 20 to 100~nm for YBCO. The deposition temperature and oxygen pressure were respectively 780~$^\circ$C and 0.25~mbar for YBCO and 800~$^\circ$C and 0.16~mbar for LSMO. For LSMO, the quality of epitaxial growth depends on the flux rate of the ablated material. We used the substrate-target distance to optimize the epitaxy of the LSMO layers. After deposition, the thin films were annealed for 10 minutes at 600 $^\circ$C in oxygen close to atmospheric pressure and subsequently cooled down at a rate of 4~$^\circ$Cmin$^{-1}$.

X-ray diffraction (XRD) measurements confirmed the epitaxial growth of the multilayers on both types of substrates. YBCO showed a slightly distorted unit cell on STO (305): the angle between the crystallographic $a$ and $c$-axes was 90.7(4)$^\circ$, resulting in a monoclinic unit cell. However, a single film on STO (305) showed an almost nominal value for $T_c$ of 90~$^\circ$C.

LSMO grows smoothly on STO~(305) substrates. Atomic force microscope (AFM) measurements on a 150~nm film showed a root mean square (rms) roughness of 2~nm and a peak-to-peak (pp) roughness of 5~nm. YBCO was much rougher with a pp roughness of 30~nm (5~nm rms) for a 100~nm film. The AFM images are shown as insets in Fig.~\ref{bilayer}. We attribute this large roughness to differences in growth rate between the YBCO $ab$ and $c$-direction. Secondly, nucleation effects are expected, since the YBCO lattice vector in the crystallographic $c$-direction is three times as large as that of STO. As a result, an integer number of YBCO unit cells will not always fit between two nucleation sites. We therefore expect a large number of antiphase boundaries in these films. When LSMO was grown on top of YBCO, the average roughness did not further increase. For bilayers, this implies that we can choose to grow a smooth LSMO/YBCO interface, by putting the LSMO underneath the YBCO layer, or a rough interface, by putting LSMO on top of YBCO, making roughness a controllable parameter in unraveling the spin-switch mechanism.

\subsection{Transport and magnetization properties \label{magnetization_prop}}

\begin{figure}
\includegraphics[scale=\schaal]{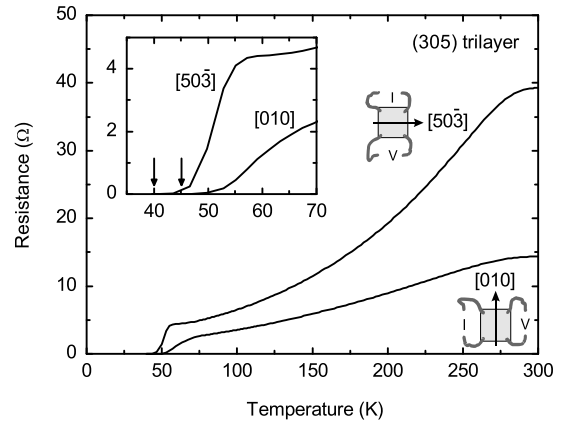}
\caption{\label{RT_trilayer} Temperature-dependence of the resistance for a (305) oriented F/S/F trilayer for two different directions of the applied current, as indicated. The layer thicknesses for the bottom F, S and top F layer are 50, 30 and 150~nm, respectively. The inset shows the behavior around $T_c$, vertical arrows indicate $T_c$.}
\end{figure}

\begin{figure}
\includegraphics[scale=\schaal]{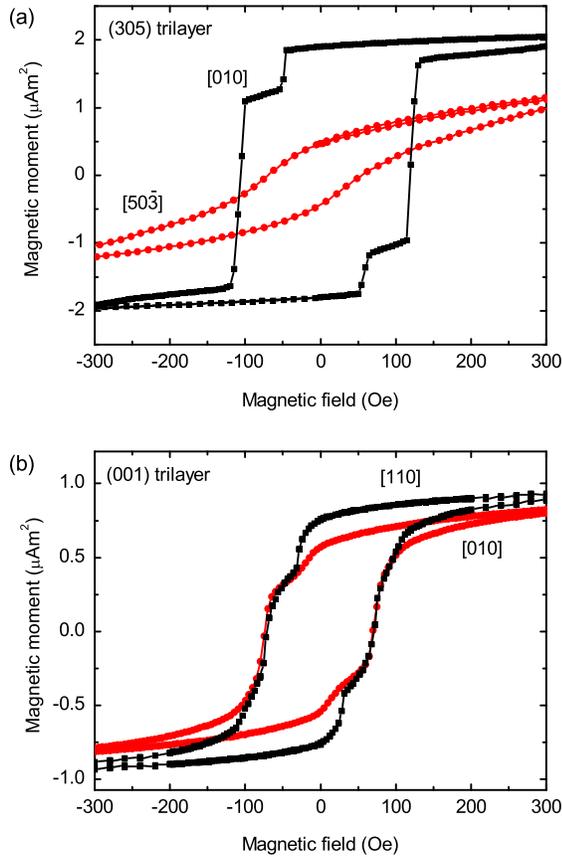}
\caption{\label{magnetization} Magnetization measurements on (305) oriented (a) and (001) oriented (b) F/S/F trilayers. The bottom and top F layers are 50 and 150~nm, respectively, the S layer is 30~nm. Measurements are taken at temperatures slightly above $T_c$ of the superconductor. The magnetic field directions are indicated in the figure. The (305) oriented trilayers show uniaxial magnetic anisotropy. The magnetization loop for the (001) oriented trilayers shows somewhat sharper features when measured along the [110] direction than along the [010] direction, in accordance with literature. \cite{mathews2005siu}  }
\end{figure}

Temperature-dependent resistance (RT) measurements on trilayers clearly showed a parallel contribution of both LSMO and YBCO. In Fig.~\ref{RT_trilayer}, RT-curves are shown that are measured for two different directions of the current in a four-point configuration with electrical connections to the corners of the trilayer. This configuration was used in all measurements. In the [010] direction the resistance has a YBCO-like linear temperature-dependence. The resistance measured in the [50$\bar{3}$] direction is larger and has the bell-shape that is typical for LSMO, indicating that the YBCO resistance is higher in this direction. We attribute this to the $c$-axis transport component, which is present for this direction. In addition, a contribution of the antiphase boundaries can be expected predominantly in this direction. The thinnest YBCO films in bi- and trilayers exhibited a reduced $T_c$, probably related to strain effects. In some structures we found two values for $T_c$ depending on the direction of measurement. Thus, a superconducting path between the current electrodes in the [010] direction could be formed at a higher temperature than in the [50$\bar{3}$] direction. By using a zero resistance criterion for $T_c$,  we found  45~K (40~K in the [50$\bar{3}$] direction), for thicknesses of 30~nm, which decreased to 20~K (both directions) for 20~nm films.

Magnetization measurements were performed using a vibrating sample magnetometer (VSM) mounted in the same system in which the transport measurements were taken. In one occasion, a SQUID magnetometer was used. Small field offsets (less than 20~Oe) observed in the VSM were absent in the SQUID magnetometer. Our thin films showed slightly reduced Curie temperatures in the range of 320 to 350~K. Hysteresis loops with the field oriented along the [010] direction and the [50$\bar{3}$] direction are presented in Fig.~\ref{magnetization}a for an F/S/F trilayer with bottom and top layers of 50 and 150~nm, respectively and a YBCO thickness of 30~nm. The contributions of the two individual LSMO layers are clearly visible and sharp magnetization switching is observed when the field is applied in the [010] easy direction. Since the magnetic anisotropy of LSMO is sensitive to strain and uniaxial strain was found to induce uniaxial magnetic anisotropy, \cite{suzuki1997rsm} we expect uniaxial magnetic anisotropy for LSMO on STO (305) as well. Indeed, the [50$\bar{3}$] direction is clearly not an easy axis. We tried to fit both curves using the Stoner-Wohlfarth model \cite{stoner1948} for a single domain ferromagnet, but could not find a satisfactory fit using a single set of parameters. The (001) oriented trilayers are expected to show biaxial magnetic anisotropy at low temperatures. \cite{mathews2005siu} Although the difference is small, the magnetization loop measured along the [110] easy direction (measured in the SQUID magnetometer) as shown in Fig.~\ref{magnetization}b shows sharper features and larger saturation magnetization than the one measured along the [010] hard direction. Although two coercive fields are observed for both directions, the switching is less sharp than for the (305) oriented trilayer and the antiparallel (AP) state is poorly defined. We conclude that this is due to the biaxial magnetic anisotropy.

\section{Results and discussion \label{results}}
\subsection{Resistance switching in F/S bilayers}

\begin{figure}
\includegraphics[scale=\schaal]{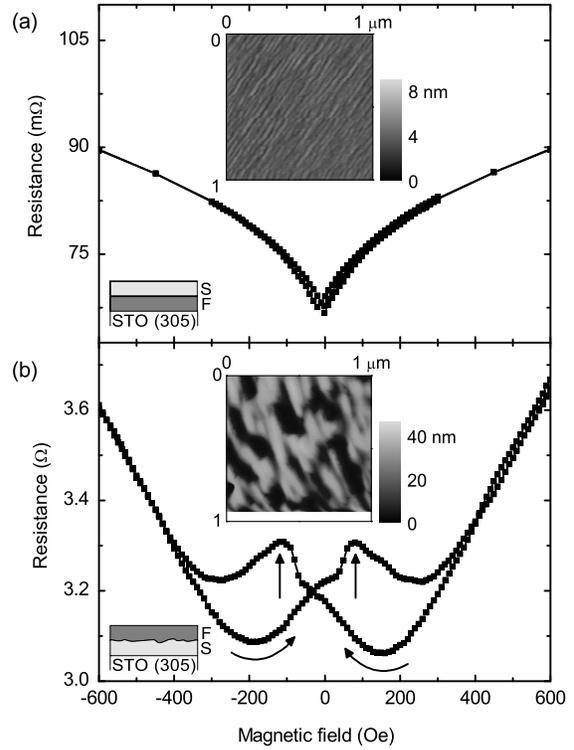}
\caption{\label{bilayer} Resistance measurements at 61~K in the superconducting transition as a function of magnetic field on (a) an STO(305)/LSMO/YBCO bilayer (150/30~nm) and (b) an STO(305)/YBCO/LSMO bilayer (30/150~nm) measured in a current-in-plane configuration. The magnetic field is applied along the [010] easy axis. The sweep direction is indicated by arrows, the vertical arrows indicate the coercive field of the ferromagnetic layer. The inset in (a) shows an atomic force microscope (AFM) image obtained on a 150~nm single LSMO film, which is much smoother than a 100~nm single YBCO film, as shown in (b). }
\end{figure}

We have grown bilayers on STO (305) both with the LSMO underneath YBCO (F/S) and with the LSMO on top (S/F). In both structures, the YBCO thickness is 30~nm and the LSMO thickness 150~nm. Both structures show a reduced $T_c$ of 60~K. The resistance as a function of magnetic field is measured in the superconducting transition (at 61~K) using a current-in-plane (CIP) technique. Magnetic fields are applied along the easy axis. In the STO(305)/F/S structure, which has a smooth LSMO layer, the observed hysteresis is the largest for temperatures above $T_c$. Even here, it is smaller than 0.2~\% and is a direct result of the butterfly-shaped magnetoresistance of the LSMO layer. The magnetoresistance in the superconducting transition at 61~K is shown in Fig.~\ref{bilayer}a. When the order of the layers is reversed, which yields a rougher interface, a large hysteresis in the superconducting transition appears, which is too large to arise from the LSMO magnetoresistance. A typical result is depicted in Fig.~\ref{bilayer}b. Starting from large negative fields, the resistance shows a parabolic dependence on the field with a minimum around $-$200~Oe. Then, reaching the positive coercive field of ~80~Oe, indicated by a vertical arrow, a peak structure can be observed in the magnetoresistance. Above 200~Oe, the resistance starts following the parabolic dependence again, however now displaced over the horizontal axis by a value of approximately 350~Oe. Since there is only one ferromagnetic layer we cannot analyze our results in terms of the relative orientation of ferromagnetic layers ruling out the spin-switch effect as a cause for the observed shift. Similarly, explanations using spin imbalance or increased quasiparticle densities fail for bilayers, since in these models there is no dependence on the direction of the spins. In fact, the observation of hysteresis effects in bilayers strongly points at an influence of the magnetization direction of the layer and its relative direction to the applied magnetic field. One can think of the total magnetic field, given by the contributions of the applied field and the stray fields of the ferromagnetic layer, as the main parameter determining the resistance of the bilayer. The peak structure around the coercive field is then most likely caused by stray fields at domain walls, due to the reorientation of magnetic domains. The larger S/F surface roughness of the STO(305)/S/F compared to the STO(305)/F/S bilayer might be expected to increase stray field effects.  \cite{schrag2000nop}  The larger hysteresis observed in the STO(305)/S/F structures confirms this picture, in agreement with ref. \onlinecite{stamopoulos2007sfb}.

\subsection{Resistance switching in F/S/F trilayers}
\begin{figure}
\includegraphics[scale=\schaal]{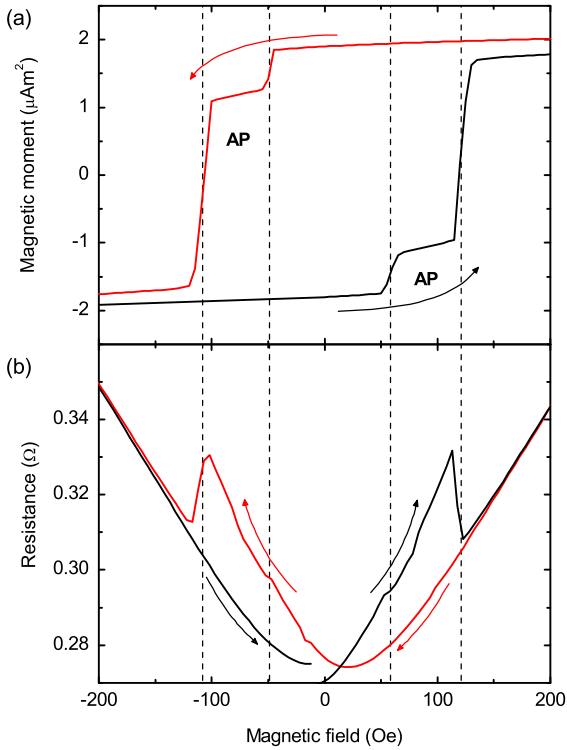}
\caption{\label{trilayer} (a) Magnetization of a (305) F/S/F trilayer (50/30/150~nm) as measured with a vibrating sample magnetometer (VSM) at 40~K. The dashed lines correspond to the coercive fields of the top and bottom layer. The highest coercive field is from the thicker top layer. The field range where the magnetization direction of the two layers is antiparallel (AP) is indicated. (b) Magnetization induced resistance switching effects in the superconducting transition (44~K). The apparent discontinuity at zero field is due to a small and smooth temperature drift in the system. Arrows denote the field sweep direction. }
\end{figure}

In addition to bilayers, we observe clear switching effects in trilayers. In Fig.~\ref{trilayer}, the magnetization curve of a (305) oriented F/S/F trilayer together with the field dependence of the resistance of the trilayer is presented. The layer thicknesses are 50, 30 and 150~nm for the bottom F, S and top F layer, respectively. The $T_c$ of the trilayer was 40~K and the measurement is performed at 44~K. When the bottom LSMO layer switches, the trilayer resistance shows a small downward deviation from the parabolic curve. A large resistance drop occurs upon switching the thicker and rougher top layer. If the resistance switching effects resulted from switching from P to AP states, an \emph{increase} in resistance of equal magnitude would be expected at the lowest coercive field. In addition, in the region around zero field, between the lowest positive and negative coercive fields, the system would be in the same P state and the curves measured in increasing magnetic field and decreasing field would have to overlap. The observed switching behavior can thus not be attributed to switching from P to AP states, but rather arises from the switching of the individual layers. It is interesting that we can observe a small resistance change as a result of the switching of the smooth F bottom layer, while we cannot see it in an STO(305)/F/S bilayer. Apparently, stray fields more easily penetrate the superconductor in trilayers than in bilayers. Similar behavior was recently observed in ref.~\onlinecite{stamopoulos2007sfb}, were it was attributed to a magnetostatic coupling of the ferromagnetic layers. 

Before discussing the data further in terms of stray fields, we would first like to discuss whether a superconducting spin-switch effect could be detectable in our system given the thickness of the superconductor being several times the coherence length of YBCO, which is about 2-3~nm in the $ab$-plane. In the original picture by Tagirov \cite{tagirov1999lfs}, the superconducting spin-switch effect depends on the parameter $(\xi_s/d_s)^2$, in which $d_s$ is the thickness of the superconducting layer and $\xi_s = \sqrt{\hbar D_s/2 \pi k_B T_c}$, $D_s$ being the diffusion constant in the superconductor, and $\hbar$ and $k_B$ are the Planck and the Boltzmann constant, respectively. The Ginzburg-Landau coherence length $\xi_{GL}$ at 0~K is approximately equal to $\xi_s$: $\xi_s = 2\xi_{GL}(0)/\pi$. \cite{Radovic1988ucf} Although the $T_c$-shift due to the spin-switch effect could be numerically calculated explicitly, we can safely conclude from the small value of $(\xi_s/d_s)^2$ that it would be small. In Ref.~\onlinecite{giazotto2006car}, a magnetoresistance effect resulting from crossed Andreev reflection processes is predicted up to approximately 10 times the coherence length. This approaches our film thicknesses, but it should be taken into account that the electrons traversing the superconductor on the $ab$-planes will experience a film thickness of 60~nm due to the 31$^\circ$ angle of the planes with respect to the sample surface. On the other hand, if the (inverse) spin-switch originates from the injection of spin-polarized electrons, the characteristic length scale is set by the spin-diffusion length in YBCO, which might well be larger than our film thickness.\cite{soltan2004fsb,nemes2008ois}

\subsection{Penetrating field model}
\begin{figure}
\includegraphics[width=.8\columnwidth]{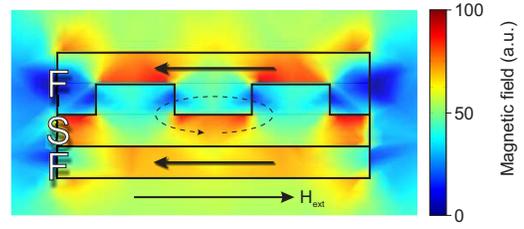}
\caption{\label{simulation} Simulated field distribution in an F/S/F trilayer with roughness. Arrows denote the field and magnetization directions. Roughness increases the field in the superconductor. At the thinnest parts of the superconductor, the stray fields are locally opposite to the magnetization direction. The situation as depicted exists when the system has been saturated in a strong negative field (pointing to the left), after which the field has been set to positive, but smaller than the lowest switching field. }
\end{figure}

\begin{figure}
\includegraphics[scale=\schaal]{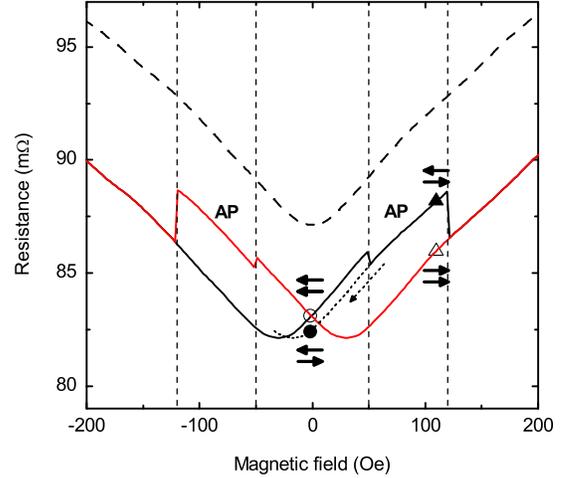}
\caption{\label{model} Reconstruction of the trilayer magnetic field dependence (solid black (red) line for increasing (decreasing) magnetic field) starting from the field dependence of a single YBCO layer in the superconducting transition (dashed curve, arbitrary offset). The vertical dotted lines denote the coercive fields of the ferromagnetic layers. The open circle and triangle denote parallel states at different field values at which antiparallel states can be prepared as well (filled symbols). Horizontal arrows represent the magnetization state of the F layers; arrows to the right (left) indicate magnetization in the positive (negative) direction.}
\end{figure}

We have shown above that the resistance switching effect in trilayers is larger when the top layer switches than when the bottom layer switches. The difference seems to be too large to arise solely from the different thicknesses of the top and bottom layer. We have already seen for the bilayers that roughness can increase the stray fields from the ferromangetic layers. If the magnetization would be perfectly homogeneous and in-plane, the field induced in the superconductor due to the magnetization of the F layers, would be very small and in fact only non-zero due to the finite size of the layers. This is the reason that in bilayers switching effects are absent when the F layer is the smooth bottom layer. To substantiate the effects of roughness further, we have carried out finite element simulations on a trilayer with one rough and one smooth F layer. Indeed, a substantial field is predicted to be induced in the superconductor, see Fig.~\ref{simulation}. In the simulation, we neglect screening effects in the superconductor, which in practice will be small, since the temperature is above $T_c$. The essential point is that in parts where the superconductor is thin (which contribute the most to the resistance), the induced field will  be \emph{opposite} to the magnetization of the layer, and can be either parallel or antiparallel to the applied field, depending on the preparation of the system. We can therefore write for the total field $B_{tot}$ in the superconductor
\begin{equation}
B_{tot} = \mu_0 \left(H_{ext} - \alpha_1  M_1 - \alpha_2  M_2\right) ,
\label{field_model}
\end{equation}
where $H_{ext}$ is the externally applied field and $\alpha_{1,2}$ are positive constants, relating the magnetization in the layers 1 and 2 to the induced field in the superconductor. It will be clear that $\alpha$ is larger for the rougher layer. Now we can combine this with the field dependence of YBCO in the absence of F layers, which is given in Fig.~\ref{model} by the dashed line. At a large positive field, the resistance will be lower than for the bare YBCO, due to the stray fields induced by the roughness, which are antiparallel to $H_{ext}$. Upon lowering the field the curve goes through a minimum at positive $H_{ext}$ because of the cancellation of external and stray fields. Further lowering yields a resistance increase because now the external and stray fields point in the same direction. At the coercive fields of the F layers 1 and 2, the curve then shift down, because the magnetization and therefore the stray fields switch and become again antiparallel. The switching of the ferromagnetic layers lead to thus to lateral shifts of the dashed curve at the coercive fields. If we take the coercive fields 50 and 120~Oe and use $\mu_0 \alpha_1 M_1 = 5$~Oe and $\mu_0 \alpha_2 M_2 = 25$~Oe, we get the curve represented by the solid line. This would correspond to values for $\alpha_{1,2}$ of 0.2~\% and 1~\%, respectively. In the light of the previously suggested superconducting spin-switch models, it is surprising that such a simple model can reproduce the observed behavior so well.

\begin{figure}
\includegraphics[scale=\schaal]{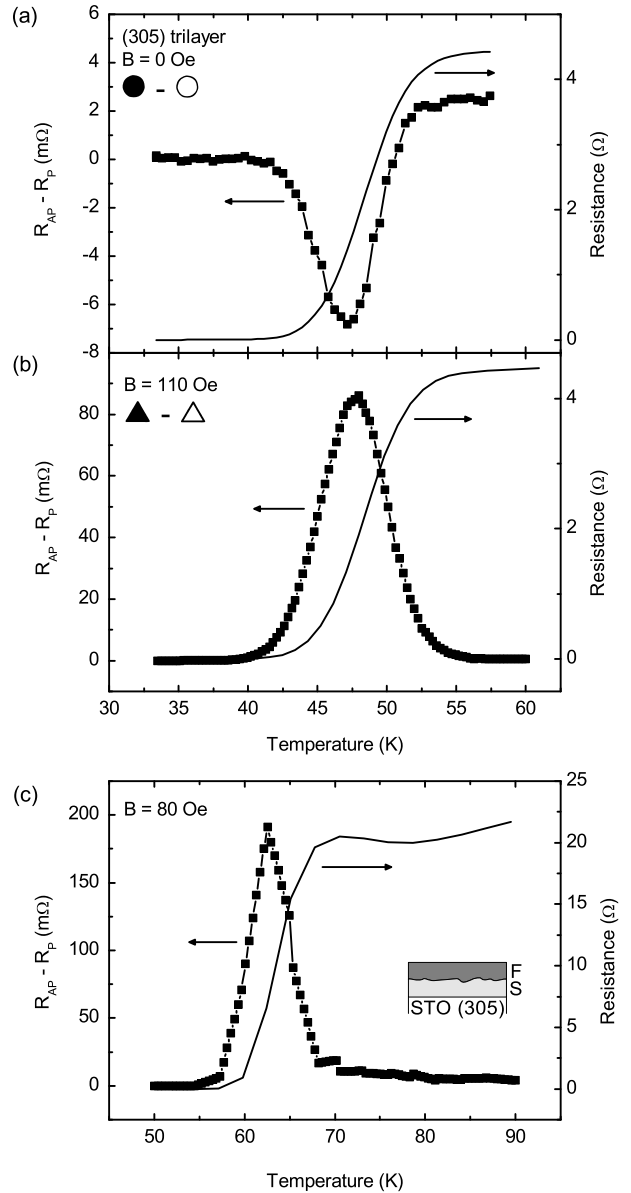}
\caption{\label{RT_switching} (a,b) Resistance differences (line and symbols) between antiparallel and parallel states for a (305) oriented F/S/F trilayer (50/30/150~nm). The symbols correspond to the symbols used in Fig.~\ref{model}. The temperature dependence of the resistance itself is indicated by the solid line (corresponding to the scale on the right). The resistance difference between the antiparallel state and the parallel state is opposite in sign and different in size for two different field values, which is difficult to account for within the spin-switch model but has a clear origin in the stray fields from the individual ferromagnetic layers, penetrating the superconductor. (c) In an S/F \emph{bilayer} (30/150~nm), at a finite field value below the coercive field, the switching of the ferromagnetic layer yields a comparable signal, supporting the idea that stray fields play an important role in these structures.}
\end{figure}

To further substantiate this result, we prepared the system to be in the states as indicated by the circles and triangles in Fig.~\ref{model} and looked at the temperature dependence of the resistance difference between the open and filled symbols. Thus, we investigated the pure effect of the switching of the top or bottom layer on the resistance. It is clear from Fig.~\ref{RT_switching} that we only see resistance differences around the superconducting transition. This is due to the fact that the magnetoresistance of YBCO above $T_c$ is small and below $T_c$, large fields are required to suppress superconductivity. Note that an increase of resistance could be interpreted as a decrease of $T_c$. At zero field, the difference between the AP and P state is small, which is due to the fact that it is the smooth bottom layer that is switched between the measurements. The signal is negative, which is clear from inspection of Fig.~\ref{model} since we are probing the difference between the filled and open circle. When we now compare this to the effect of switching the upper layer again parallel to the bottom layer, i.e. taking the difference between the open and filled triangle, we find a much larger signal of positive sign. It is interesting to see that we can mimic this behavior in a bilayer by measuring in a finite field (below the coercive field) with the magnetization AP and P with respect to the field. In Fig.~\ref{RT_switching}c we find a resistance switching effect that has similar sign and magnitude as found in the trilayer. 

We have also studied the effect of inhomogeneous magnetization in the layers by either applying a demagnetization procedure or by applying fields perpendicular to the sample. We find in both cases an increase in the resistance, which we attribute to the increased contribution of magnetic stray fields as was also found for F/S/F triple layers with perpendicular magnetic anisotropy by Singh \emph{et al.} \cite{singh2007msp} 

\subsection{Switching in (001) oriented F/S/F trilayers}

\begin{figure*}
\includegraphics[scale=\schaal]{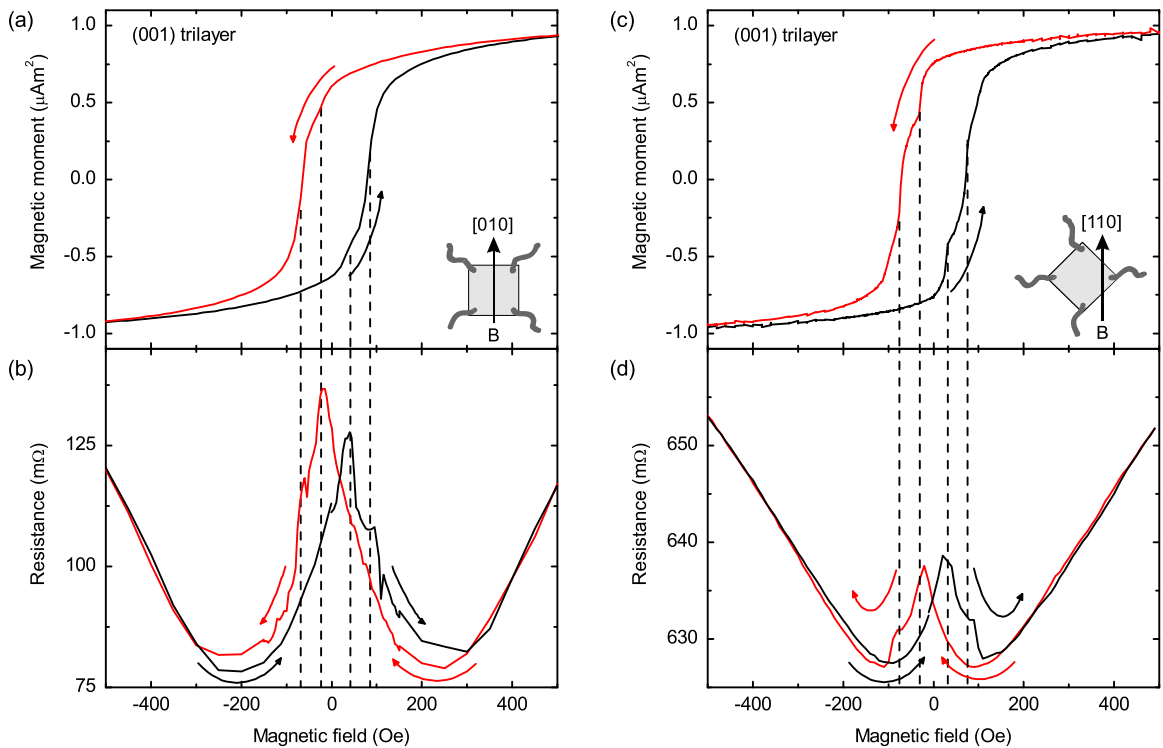}
\caption{\label{switching_caxis} Magnetization (a \& c) and resistance (b \& d) measurements of a (001) oriented F/S/F trilayer (50/30/150~nm) at 61~K in the superconducting transition for different magnetic field orientations as indicated. At the coercive field values, indicated by vertical dashed lines, resistance switching is observed. When the field is applied in the [010] direction, an increase in the resistance is observed between $-$200 and 200~Oe. This increase arises from in-plane domain reorientation effects which correspond to the rounding of the magnetization curve. When the field is applied in the [110] easy direction the rounding decreases, resulting in a reduced resistance increase.  }
\end{figure*}

We have also fabricated a (001) oriented F/S/F trilayer, using the same layer thicknesses as were used for the (305) trilayer. The trilayer showed a $T_c$ of 60~K. In section \ref{magnetization_prop}, we have seen that for the (001) oriented structures the magnetization switching is less well-defined than for the (305) oriented structures. Still, we observe resistance switching effects near the coercive fields, indicated by dashed lines in Fig.~\ref{switching_caxis}. Our data on (001) oriented structures are similar to data published in the literature. \cite{pena2005gmf,nemes2008ois} Measurements are taken at 61~K. The resistance switching effects are superimposed on a background dip which will be discussed below in section \ref{highfield}. When the field is applied along the [010] direction, an increase in the resistance is observed between $-$200 and 200~Oe, in the regime where the hysteresis loop of the magnetization starts to open.  Switching is not as sharp as in the case of the (305) trilayers, and we propose that the increase of the resistance here is due to non-homogeneous magnetization as a result of in-plane domain reorientation. Important to note is that at both switching field the resistance appears to go down rather than up, again suggesting that for each layer the direction with respect to the applied field is more important than their relative orientations. When the field is applied along the [110] easy axis, the magnetization loop is sharper and domain reorientation effects play less a role. The effect on the resistance is clear; the increase in resistance between $-$200 and 200~Oe reduces dramatically. Notice that the smoother growth of YBCO on STO (001) diminishes the difference in roughness between the top and bottom layer and the roughness of both interfaces will be comparable to the bottom interface in the (305) structures. We can thus only explain the strong resistance change from domain effects, which certainly are present, as the magnetization loop is still rounded. This probably underlies dissimilarities between the data obtained on (305) and (001) oriented trilayers. 

Let us now compare the relative magnitude of the resistance switching effect for the (001) structures with that for the (305) structures. We adopt the definition $\Delta R = (R_{max}-R_{min})/R_{nor}$, \cite{stamopoulos2007sfb} in which $R_{min}$ and $R_{max}$ are the resistance minimum and the maxium induced by the switching and $R_{nor}$ the resistance of the trilayer in the normal state. We find $\Delta R =$~0.7~\% for the (001) trilayer when the field is applied in the [010] direction and 0.2~\% when applied in the [110] direction. For the (305) trilayer the individual contributions of both layers are clearly visible and we find 0.4~\% when the top layer is switched and we estimate 0.04~\% for the bottom layer. We thus obtain that the magnitude of the resistance switching in the (305) structure is relatively large, given the sharp magnetization switching, which we attribute to the roughness of the corresponding interface. The much smoother bottom interface shows indeed a smaller switching effect than the (001) oriented structures. 

\subsection{Switchable coupling of F layers}
\begin{figure}
\includegraphics[scale=\schaal]{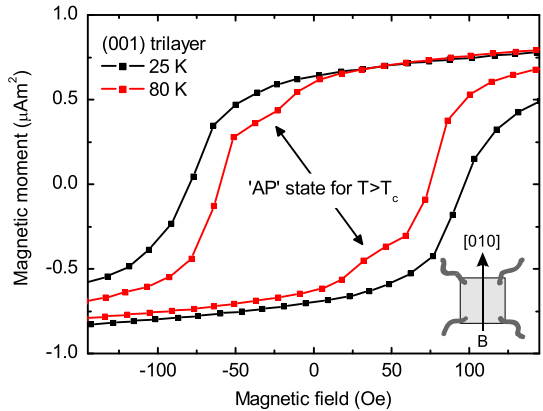}
\caption{\label{vanishing_APstate} Upon decreasing the temperature below $T_c$ of the superconductor in a (001) F/S/F trilayer (50/30/150~nm), we observe the loss of the 'AP' state due to a change in the mutual influence of the layers. This observation provides further proof that the F layers feel each others magnetic field and therefore, field effects on the superconductor can not be neglected. 'AP' is put between quotation marks here, since due to the biaxial magnetic anisotropy, it is questionable whether this state is truly antiparallel. }
\end{figure}

We have made another observation that indicates the importance of the magnetic field penetrating the superconductor in this particular kind of structures. In Fig.~\ref{vanishing_APstate} we show magnetization loops of a (001) oriented F/S/F trilayer both above $T_c$ at 80~K and well below $T_c$ at 25~K. Although, as stated above, structures with this orientation do not show single domain magnetization switching behavior, we can observe a step-like magnetization curve well above $T_c$, arising from two independent coercive fields. When the temperature is lowered to below $T_c$, this two-step behavior disappears and the coercive fields seem to merge. This behavior is likely due to the sudden change in screening behavior of the S layer. The interplay between magnetic domain structures and vortices were studied in ref.'s~\onlinecite{laviano2005cvm,laviano2007ibv,gozzelino2006qmo}. It is well-known that superconductivity in S/F hybrid structures can modify the magnetization state. \cite{monton2007mbs, wu2007eem, monton2008msm} While it is difficult here to identify exactly the mechanism leading to the observed coupling of the ferromagnetic layers through superconductivity, it is clear from the measurement that magnetic interactions between the F layers through the superconductor take place, which stresses the importance of stray fields in these structures.

\subsection{High-field behavior of the magnetoresistance \label{highfield}}

\begin{figure}
\includegraphics[scale=\schaal]{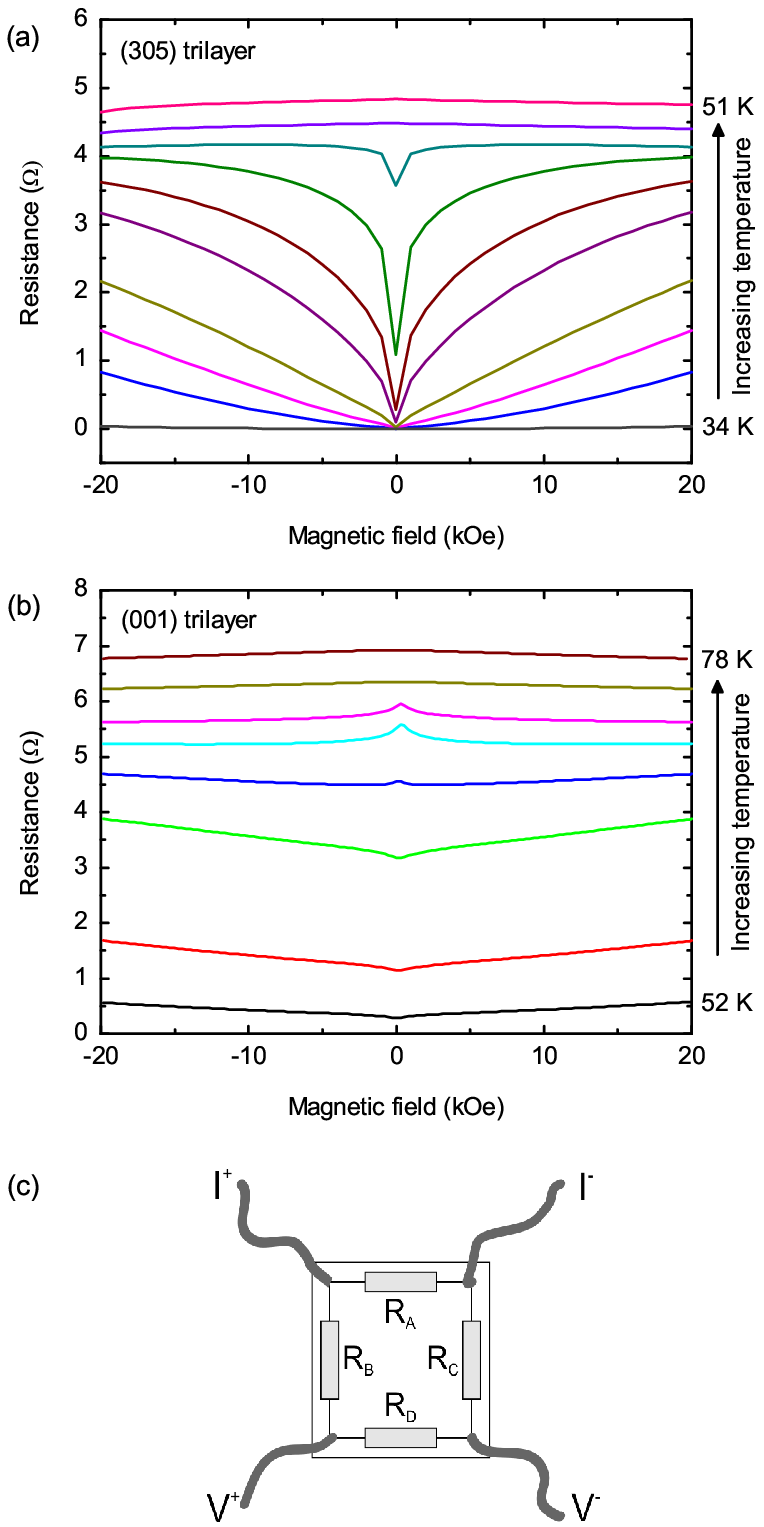}
\caption{\label{high_field} High-field magnetoresistance behavior for various temperatures from just below $T_c$ to just above $T_c$ for (a) (305) and (b) (001) oriented trilayers. In both trilayers, the bottom F, S and top F layer are 50, 30 and 150~nm, respectively. In the (305) structure we find a dip, reflecting the magnetoresistance of the YBCO in the superconducting transition. The (001) structures show a crossover from a peak to a dip centered around zero field. (c) Resistor network representing a simplified scheme of the sample resistance. When $R_B$ and $R_C$ decrease the measured resistance increases. This effect might explain the peak to dip crossover observed in (b). }
\end{figure}

Finally we would like to discuss the high-field behavior of the F/S/F trilayers. In Ref. \onlinecite{pena2005gmf}, peaks in the magnetoresistance, centered at zero field, were attributed to spin imbalance due to the injection of spin-polarized carriers in a fashion that resembles the giant magnetoresistance (GMR) effect. In Fig.~\ref{high_field}a the high-field dependence of a (305) oriented trilayer at temperatures in the range from 34 to 51~K is displayed. We observe a dip, rather than a peak, which directly reflects the magnetic field dependence of the YBCO in the superconducting transition. Note that the switching effects that have been discussed in the previous sections take place at the bottom of the dip. In a (001) oriented trilayer, however, we observe the crossover from a peak to a dip depending on the temperature, see Fig.~\ref{high_field}b. We propose a straightforward explanation for this crossover. Especially around $T_c$, small inhomogeneities in the film can lead to large resistance variations over the sample. For example, a small variation of $T_c$ over the sample can lead to a considerable resistance variation over the sample. With the help of a simplified resistor network in Fig.~\ref{high_field}c it is easy to see that when a current, $I_{tot}$, is passed through the current contacts $I^+$ and $I^-$ the voltage over the voltage contacts ($V_D$) will be given by $I_{tot} R_A R_D/(R_A + R_B + R_C + R_D)$. This means that when the resistances $R_B$ and $R_C$ \emph{decrease}, the resistance we measure (i.e. $V_D/I_{tot}$) \emph{increases}. Indeed, in (305) oriented trilayers, where superconducting paths are achieved at higher temperatures in the [010] direction than in the [50$\bar{3}$] direction, the superconducting transition in one direction is sometimes accompanied by a resistance increase in the other direction. In Fig.~\ref{RT_trilayer} a weak signature of this effect can be seen. In a similar way, if a superconducting path is achieved in the direction perpendicular to the one in which the measurement is performed (in Fig.~\ref{high_field}c for example $R_B$), this will generate a magnetoresistance with a dip, which now appears as a peak in the actual measurement. For lower temperatures, the dip in the initial superconducting path becomes weaker, but a direct superconducting connection between the voltage contacts will appear, resulting in the recovery of a dip.

\section{Conclusion\label{conclusion}}

We have searched for the superconducting spin-switch effect in F/S/F LSMO/YBCO bi- and trilayers that were optimized for the effect by making the contact between the materials partly in YBCO's crystallographic $ab$-plane. Although we find sharp magnetization switching behavior in these structures, with a well-defined antiparallel state, we do not observe any signature of a spin-switch effect. Instead, our data provide compelling evidence that the observed resistance switching effects are caused by magnetic stray fields from the ferromagnetic layers, and that also interface roughness can play a role in the observed effects. In the case of the sharply switching (305) oriented structures, we find that we can explain the data by taking such roughness into account explicitly. In (001) oriented structures, we have shown that domain reorientation effects have a strong contribution. Moreover, the same description allows to explain data taken on bilayers with either rough or smooth interfaces. The results may be a warning sign that magnetic field effects, although often not considered to play a role in this kind of structures, might be important after all.

\begin{acknowledgments}
This work is part of the research program of the Foundation for Fundamental Research on Matter (FOM), financially supported by the Netherlands Organization for Scientific Research (NWO), and the NanoNed program.
\end{acknowledgments}

\bibliography{LSMO_305}

\end{document}